\documentclass[12pt]{article}
\usepackage{epsfig}
\newdimen\unit
\def\point#1#2#3{\vbox to0pt{\kern-#2\unit
  \hbox{\kern#1\unit$#3$}\vss}
 \nointerlineskip}
\newcommand{\nc}{\newcommand}

\def\eqn#1{eq.~(\ref{#1})}

\nc{\fig}[1]{Fig.~\ref{fig:#1}}
\nc{\figs}[2]{Figs.~\ref{fig:#1} and \ref{fig:#2}}
\renewcommand{\slash}[1]{/\kern-7pt #1}
\nc{\beq}{\begin{equation}}
\nc{\eeq}{\end{equation}}
\nc{\bea}{\begin{eqnarray}}
\nc{\eea}{\end{eqnarray}}
\nc{\beas}{\begin{eqnarray*}}
\nc{\eeas}{\end{eqnarray*}}
\nc{\nn}{\nonumber}
\nc{\gsim}{\raisebox{.2em}{$\rlap{\raisebox{-.5em}{$\;\sim$}}>\,$}}
\nc{\lsim}{\raisebox{.2em}{$\rlap{\raisebox{-.5em}{$\;\sim$}}<\,$}}
\nc{\eps}{\epsilon}

\def\eps{\epsilon}
\def\qb{{\bar q}}

\def\spa#1.#2{\left\langle#1 \hskip .15 mm #2\right\rangle}
\def\spb#1.#2{\left[#1 \hskip .15 mm #2\right]}
\def\spaa#1.#2.#3{\langle\mskip-1mu{#1} 
                  | #2 | {#3}\mskip-1mu\rangle}
\def\spab#1.#2.#3{\langle\mskip-1mu{#1} 
                  | #2 | {#3}\mskip-1mu\rangle}
\def\cg{c_\Gamma}

  \newcommand{\ccaption}[2]{
    \begin{center}
    \parbox{0.85\textwidth}{
      \bigskip
      \caption[#1]{\small{#2}}
      }
    \end{center}
    }
\newbox\charbox
\newbox\slabox
\def\s#1{{      
        \setbox\charbox=\hbox{$#1$}
        \setbox\slabox=\hbox{$/$}
        \dimen\charbox=\ht\slabox
        \advance\dimen\charbox by -\dp\slabox
        \advance\dimen\charbox by -\ht\charbox
        \advance\dimen\charbox by \dp\charbox
        \divide\dimen\charbox by 2
        \raise-\dimen\charbox\hbox to \wd\charbox{\hss/\hss}
        \llap{$#1$}
}}

\newcommand{\h}{\tilde{h}}
\newcommand{\lra}{\leftrightarrow}
\newcommand{\dk}{\Delta\kappa}

\begin{document}

\begin{titlepage}
\begin{flushright}
ETH-TH/00-04  \\
DTP/00/10 \\
February 2000 
\end{flushright}
\vskip 1.2cm
\begin{center}
\boldmath
{\Large\bf $W\gamma$ and $Z\gamma$ production at Hadron Colliders} 
\vskip .2cm

\unboldmath 
\vskip 1.2cm
{\sc D. de Florian$^a$  and  A. Signer$^b$}
\vskip .8cm
{\it $^a$ Theoretical Physics, ETH Z\"urich, Switzerland }
\vskip .3cm
{\it $^b$ Department of Physics, University of Durham,
Durham DH1 3LE, England}
\vskip 5.cm

\end{center}

\begin{abstract}
\noindent 
We present a general purpose Monte Carlo program for the calculation
of any infrared safe observable in $W\gamma$ and $Z\gamma$ production
at hadron colliders at next-to-leading order in $\alpha_s$.  We treat
the leptonic decays of the $W$ and $Z$-boson in the narrow-width
approximation, but retain all spin information via decay angle
correlations. The effect of anomalous triple gauge boson couplings is
investigated and we give the analytical expressions for the
corresponding amplitudes. Furthermore, we propose a way to study the
effect of anomalous couplings without introducing the ambiguity of
form factors.

\end{abstract}

\end{titlepage}

\section{Introduction}

The production of $W\gamma$ and $Z\gamma$ in hadronic collisions has
been studied extensively since the Born cross sections have been
computed~\cite{BMS79,Renard82}. In particular, these processes allow
to study the triple gauge boson couplings $WW\gamma$, $ZZ\gamma$ and
$Z\gamma\gamma$. The study of these couplings is mainly motivated by
the hope that some new physics may modify them. If the new physics
occurs at an energy scale well above that being probed experimentally,
it is possible to integrate it out. The result is an effective theory
which might result in non standard triple gauge boson couplings.

Both collaborations at the Tevatron have studied the production of
$W\gamma$~\cite{expWgamma} and $Z\gamma$~\cite{expZgamma} pairs. The
bounds on the anomalous couplings obtained at the Tevatron tend to be
less constraining than those obtained at LEP~\cite{expLEP}. However it
has to be kept in mind that these analyses are complementary. At the
hadron colliders a whole range in the center of mass energy is tested,
whereas at LEP the center of mass energy is fixed by the collider.
Furthermore, with Run~II the expected number of events at the Tevatron
increases substantially. Assuming a data sample of 2~fb$^{-1}$, more
than 3000 $W\gamma\to\ell\nu\gamma$ events and 700
$Z\gamma\to\ell\ell\gamma$ events are expected for each
experiment~\cite{Diehl98}. Of course, the expected number of events
is even bigger for the LHC.

Anomalous triple gauge boson couplings lead to deviations from
Standard Model predictions. Obviously, observables for which these
deviations are enhanced offer better chances to find new physics or
get tighter constraints on anomalous couplings. There are basically
two classes of such observables. Either we consider observables which
are strongly suppressed in the Standard Model or observables with
large transverse momentum (or center of mass energy). In both cases,
the inclusion of next-to-leading order (NLO) QCD corrections is mandatory.

A prominent example of an observable that is suppressed in the
Standard Model is the so called radiation zero for $W\gamma$
production. At leading order (LO) there exist some kinematic
configurations for which the amplitude vanishes~\cite{BMS79}. This is
manifest in some observables as a dip in the rapidity distributions.
Since anomalous coupling contributions fill in the dips, there seemed
to be excellent prospects to obtain accurate limits for them from
experimental data.  Unfortunately, next-to-leading order QCD 
corrections strongly affect the LO analysis. They have the same effect
as the anomalous coupling contributions. The dips are filled in,
making the extraction of anomalous couplings quite more difficult.

For processes with large transverse momentum or center of mass energy,
the NLO corrections are particularly large. This is due to the fact
that the cross sections in these cases get large contributions from
gluon induced partonic subprocesses, which only enter in a
next-to-leading order description of the cross section. Thus, even
though the anomalous contributions are enhanced in these regions, a
calculation at NLO in $\alpha_s$ is required to reliably exclude (or
establish) physics beyond the Standard Model.

The relevance of NLO corrections was first shown for the production of
real (spin-summed) $W$ and $Z$ bosons with Standard Model couplings
and without considering lepton decays and spin correlations
\cite{STvN89,OhnemusWZ93}. These calculations were later extended, in
order to include the leptonic decays and anomalous couplings
\cite{BHOW93,OhnemusZ95,BHOZ98}. However, the full one-loop amplitudes
including leptonic decays became available only very recently
\cite{DKS98}. Therefore, refs.~\cite{BHOW93,OhnemusZ95,BHOZ98}
included decay correlations everywhere except for the finite part of
the virtual contributions.

In this paper we present order $\alpha_s$ results for the production
of $W\gamma$ and $Z\gamma$ in hadronic collisions, including the
{\it full} leptonic correlations. We work in the narrow-width
approximation, where only `single-resonant' Feynman diagrams have to
be considered. The simplicity of the helicity method allows to take
into account anomalous couplings as well and present for the first
time analytical expressions for the corresponding amplitudes.

For the case of $WW, \ ZZ$ and $WZ$ production at hadron colliders,
some results beyond the narrow-width approximation are known. The
narrow-width approximation requires only the calculation of
`doubly-resonant' Feynman diagrams. However, for these processes, also
the amplitudes including `single-resonant' diagrams have been computed
and implemented into a Monte Carlo program~\cite{CE99}.

To perform the phase space integration we use the subtraction method
discussed in ref.~\cite{FKS}. This allows for a straightforward
implementation of the one-loop $q\bar{q}' \rightarrow V\gamma
\rightarrow \ell \ell' \gamma$ and bremsstrahlung $q\bar{q}'
\rightarrow g V\gamma \rightarrow g \ell \ell' \gamma$ amplitudes,
presented in ref.~\cite{DKS98} $(V\in\{Z,W\})$. The constructed Monte
Carlo code allows the computation of any infrared-safe
observable\footnote{The corresponding Fortran codes are available upon
request}.

A brief overview of the calculation is given in section 2, were we
summarize the input parameters used and the cuts implemented to obtain
our phenomenological results. In section 3 we first present some
benchmark cross section numbers for both $W\gamma$ and $Z\gamma$
production at the LHC and study the typical scale dependence of some
observables at NLO in the Standard Model. Since many distributions
have been studied in the past, we refrain from doing a detailed
analysis. However, as soon as more precise data becomes available such
an analysis can easily be done.

In section 4 we concentrate on anomalous triple gauge boson
couplings. We describe the parameterization of the triple gauge boson
vertex in terms of anomalous coupling parameters and search for the
kinematical region where its effect is amplified, namely at large
transverse momentum for both photon and leptons. We also analyze the
possibility of seeing the effect of approximate radiation zeros in the
$W\gamma$ process, i.e. by looking for 'dips' in rapidity
distributions. In order to avoid the arbitrariness introduced by form
factors, we propose to analyze the anomalous couplings as a function
of the squared partonic center of mass energy $\hat{s}$. This has been
suggested previously for the $Z\gamma$ case~\cite{GLR99}, where such
an analysis is straightforward. We extend this idea to the $W\gamma$
production. This case is more involved, since a complete
reconstruction of $\hat{s}$ is impossible, due to the appearance of a
non observed neutrino in the $W$ decay. Particularly, we present an
observable quantity which is highly correlated to $\hat{s}$ and,
therefore, allows such an analysis even for $W\gamma$ production.
Finally, in section 5 we give our conclusions and in the appendix we
present analytical expressions for the amplitudes relevant for the
processes under consideration.

\section{Formalism}

The helicity amplitudes needed for the computation of the NLO
corrections to $W\gamma$ and $Z\gamma$ production in the Standard
Model were presented in ref. \cite{DKS98}. The amplitudes relevant for
the inclusion of anomalous couplings are presented in the appendix. In
order to cancel analytically the soft and collinear singularities
coming from the bremsstrahlung and one loop parts, we have used the
version of the subtraction method presented in ref.~\cite{FKS}.  The
amplitudes are therefore implemented into a numerical Monte Carlo
style program, which allows to calculate any infrared-safe physical
quantity with arbitrary cuts.
 
Obviously, the Monte Carlo program can be used for the Tevatron and
the LHC. However, in this paper we will mainly concentrate on results
for the LHC collider, which corresponds to $pp$ scattering at
$\sqrt{s}=14$ TeV. Unless otherwise stated, the results are obtained
using the following cuts: we make a transverse momentum cut of
$p_T^{\ell}>25$~GeV for the charged leptons and the rapidity is
limited to $|\eta|<2.4$ for all detected particles. The photon
transverse momentum cut is $p_T^{\gamma}>50 (100)$~GeV for $W\gamma$
($Z\gamma$) production.  For the $W\gamma$ case we require a minimum
missing transverse momentum carried by the neutrinos $p_T^{\rm
miss}>50$~GeV. Additionally, charged leptons and the photons must be
separated in the rapidity-azimuthal angle by $\Delta R_{\ell\gamma} =
\sqrt{ (\eta_\gamma -\eta_\ell)^2 + (\phi_\gamma -\phi_\ell)^2}> 0.7$.
Moreover, since our calculation is done in the narrow-width
approximation and, therefore, ignores the radiation of photons from
the final state leptons, we apply an additional cut to suppress the
contribution from the off-resonant diagrams. For that purpose, we
require the transverse mass $M_T>90$~GeV for $W\gamma$ production and
the invariant mass of the $\ell\ell\gamma$ system $M_{\ell\ell\gamma}>
100$~GeV for the $Z\gamma$ case.
 
Finally, photons can also be significantly produced at LHC from the
{\it fragmentation} of a final state parton\footnote{This
contribution is also known in the literature as `bremsstrahlung'}.
Unfortunately, fragmentation functions of partons into photons are not
very well determined and the NLO calculation for such a contribution
is not available yet. In principle, a full NLO calculation should
include it however, since only the sum of the `direct' plus
`fragmentation' components is physically well defined at NLO (only in
the sum all collinear singularities cancel out). In order to
circumvent this problem, we include the LO component of the
fragmentation part but using NLO fragmentation distributions, where we
can factorize the final state $q\gamma$ collinear singularities.
Since the `fragmentation' component can be further suppressed
implementing certain cuts (see below) the lack of its NLO calculation
is not expected to affect the final result beyond the few percent
level.

The fragmentation contribution constitutes a background to the search
of anomalous couplings, since it does not involve any triple gauge
boson coupling.  Fortunately, there is a way to suppress its
contribution by requiring the photons to be isolated from hadrons. In
this paper we require the transverse hadronic momentum in a cone of
size $R_0=0.7$ around the photon to be smaller than a small fraction
of the transverse momentum of the photon
\begin{equation}
\sum_{\Delta R<R_0} p_T^{\rm had}< 0.15\, p_T^{\gamma}
\end{equation}
This completes the definition of our 'standard' cuts.

When indicated, we also apply a jet-veto, which means that we reject
any event where a jet of $p_T^{\rm jet}>50$~GeV and $|\eta_{\rm
jet}|<2.5$ is observed.

In our results we do not include the branching ratios of the vector
boson into leptons. They can be taken into account by simply
multiplying our final results with the corresponding branching
ratio. For both the LO and NLO results we use the latest (corrected)
set of parton distributions of MRST(cor01) \cite{MRST} and the two loop
expression (with $n_f=5$ for the typical scales of these processes)
for the strong coupling constant with $\Lambda_{\overline{\rm
MS}}(n_f=4)=300$~MeV  which corresponds to $\alpha_s(M_Z)=0.1175$. For the
fragmentation component we use the fragmentation functions from
ref.~\cite{GRVfrag}.

Since we are particularly interested in the large $p_T$ tail, which is
more sensitive to the anomalous coupling contributions, we use (unless
otherwise stated)
\begin{equation}
\mu^2=\mu^2_{\rm st}\equiv M_V^2 +\frac{1}{2} \left[
(p_T^V)^2 + (p_T^\gamma)^2
\right] .
\label{must} 
\end{equation}
as the 'standard' scale for both the factorization and renormalization
scales. 

Contributions from $b$ and $t$ quark initial states have been
neglected and, consistently, the following values have been used for
the Cabibbo-Kobayashi-Maskawa (CKM) matrix elements in the case of
$W\gamma$ production: $|V_{ud}|=|V_{cs}|=0.975$ and
$|V_{us}|=|V_{cd}|=0.222$.

The masses of the vector bosons have been set to $M_Z=91.187$~GeV and
$M_W=80.41$~GeV. We do not include any QED or electroweak corrections
but choose the coupling constants $\alpha$ and $\sin^2\theta_W$ in the
spirit of the `improved Born approximation' \cite{IBA1,IBA2}, with
$\sin^2\theta_W=0.230$. Notice that the observable is order
$\alpha^2$; within the same spirit we use the running
$\alpha=\alpha(M_Z)=1/128$ for the coupling between the vector boson
and the quarks (to effectively take into account the EW corrections)
whereas we keep $\alpha=1/137$ for the photon coupling. It is worth
noticing that this modification results in a 7\% change in the
normalization of the cross section with respect to the standard
approach of using both running coupling constants.

\section{Standard Results}

We begin the presentation of our results with some total cross section
numbers in Table~\ref{tab:XStev}, which can be useful for future
checks and for an estimate of the number of events to be observed at
the LHC. The first three results were obtained by imposing only the cut
on the transverse momentum of the photon, i.e. $p_T^\gamma>50(100)$
GeV for $W\gamma$ ($Z\gamma$) production. Apparently, the NLO
corrections as well as the fragmentation contribution are very large.
As discussed above, the relative importance of the fragmentation
contribution can be reduced substantially by applying the isolation
cut prescription.  This can be seen from the results for the total
cross section obtained after the implementation of our standard cuts,
which are also presented in Table~\ref{tab:XStev}. Unfortunately, most
of the previous publications on the subject
\cite{OhnemusWZ93,BHOW93,OhnemusZ95,BHOZ98} do not present cross
sections numbers. Nevertheless, we have compared many of the plots
shown in refs.~\cite{OhnemusWZ93,BHOW93,OhnemusZ95,BHOZ98}, specially
for the case of real (spin-summed) $W/Z \gamma$ production
\cite{OhnemusWZ93} (which is not affected by lepton
correlations). Within the precision that can be reached in such a
comparison, we found good agreement.

\begin{table}[h]
\begin{center}
\vskip0.2cm
\begin{tabular}{|c||c|c|c|} \hline 
$\sigma$ & LO$^*$ & Frag. & NLO  \\ \hline  \hline
$ W^+\gamma \,\,\, (p_T^\gamma >50$ GeV) 
  & 4.79  & 3.02 & 13.89  \\ \hline
$ W^-\gamma \,\,\, (p_T^\gamma >50$ GeV)
  & 3.08 & 3.55 & 10.15  \\ \hline
$ Z\gamma \,\,\, (p_T^\gamma >100$ GeV)
  & 1.29 & 0.412 & 2.37 \\ \hline
$ W^+\gamma $ (std. cuts) 
  &0.436   & 0.094 & 1.71  \\ \hline
$ W^-\gamma$ (std. cuts) 
  & 0.310 & 0.095 & 1.20  \\ \hline
$ Z\gamma$ (std. cuts)
  & 0.524 & 0.041 & 0.877\\ \hline
 \hline
\end{tabular}
\end{center}
\caption[dummy]{\small Cross sections in pb for $pp$ collisions at
$\sqrt{s}=14 $ TeV for $\mu = \mu_{\rm st}$. The statistical errors
are $\pm$1 within the last digit.  LO$^*$ corresponds to the direct
component only.
\label{tab:XStev}}
\end{table}

In what follows we will estimate the theoretical uncertainty of our
results by analyzing the variation of various distributions when
changing the scale $\mu$ by a factor of two in both directions $1/2 \mu_{\rm
st} \le \mu \le 2 \mu_{\rm st}$. Since many observables already have
been studied in the past~\cite{OhnemusWZ93,BHOW93,OhnemusZ95,BHOZ98}
and in order to avoid the proliferation of plots, we refrain from
presenting a detailed analysis here. We simply concentrate on a couple
of typical examples in order to give a general picture and illustrate
the importance of NLO corrections in both $W\gamma$ and $Z\gamma$
production.

\begin{figure}[h]
\centerline{ \epsfig{figure=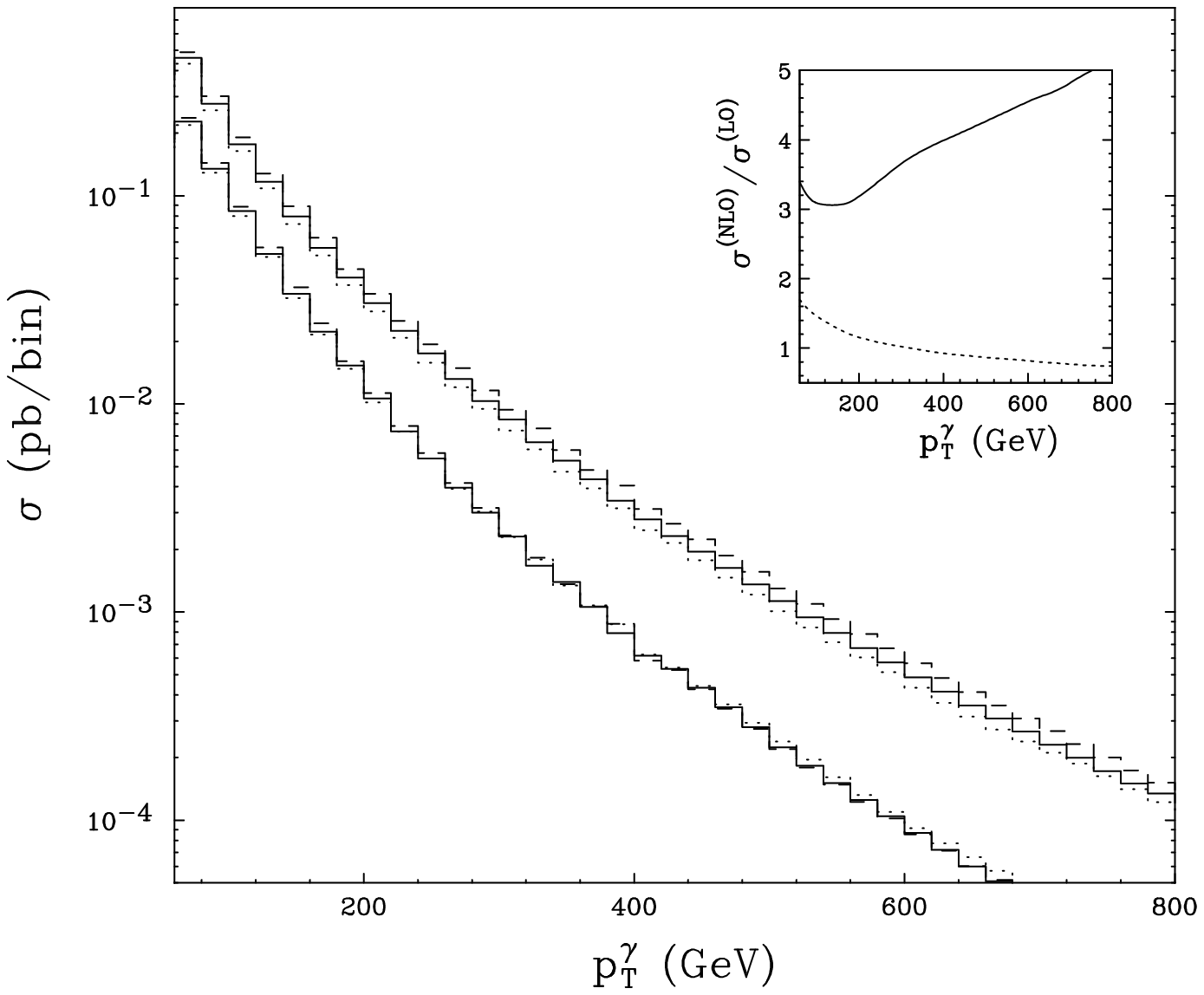,width=0.64\textwidth,clip=} }
   \ccaption{}{ \label{fig:scaleWY} Scale dependence of
   $\sigma^{(NLO)}$ without (upper curves) and with (lower curves)
   jet-veto. The scale has been varied according to $\frac{\mu_{\rm
   st}}{2} ({\rm dashes}) < \mu < 2 \mu_{\rm st} ({\rm dots})$. The
   inset plot shows the ratio $\sigma^{(NLO)}/\sigma^{(LO)}$, again
   without (solid) and with (dots) jet-veto.  }
\end{figure}

In Figure~\ref{fig:scaleWY} we show the scale dependence of the $p_T$
distribution of the photon in $W^+\gamma$ production with the standard
cuts applied (upper curves) and also with the additional requirement
of a jet-veto (lower curves). As can be observed, the scale dependence
is still large ($ ~ \pm 10 \% $) as long as only the standard cuts are
applied. However, it is considerably reduced when the jet-veto is
applied. The situation is similar to what has been observed in the
case of $WW$ production \cite{DKS99} and is caused by the suppression
of the contribution from the $qg$ initial state appearing for the
first time at NLO. Since this initial state dominates the cross
section, the NLO result behaves, regarding the scale dependence,
effectively like a LO one.

In the inset plot we present the ratio between the NLO and LO results
(with the standard scale), which remains larger than 3 and increases
with the photon transverse momentum.  This clearly shows that the LO
calculation is not even sufficient for an understanding of the shape
of the distribution, since the NLO effect goes beyond a simple change
in the normalization. As is well known \cite{OhnemusWZ93}, the
relevance of the NLO corrections for this process is mainly due to the
breaking of the radiation zero appearing at LO and to the large $qg$
initial state parton luminosity at the LHC. We also show the ratio of
the NLO jet-veto and the LO result. As expected, this ratio is closer
to 1, again due to the fact that most of the contributions coming from
the new subprocesses appearing at NLO are suppressed by the jet-veto.

It is worth mentioning that the scale dependence of the LO result
turns out to be very small. This is an artificial effect and
illustrates that a small scale dependence is by no means a guaranty
for small NLO corrections. In fact, there is no renormalization scale
dependence at all at LO. The only scale dependence comes from the
factorization scale dependence of the parton distribution
functions. Furthermore, we would like to mention that the situation
concerning the scale dependence is slightly more favorable at the
Tevatron. This is simply due to the fact that the gluon initiated
process is less important.

\bigskip

\begin{figure}[h]
\centerline{ \epsfig{figure=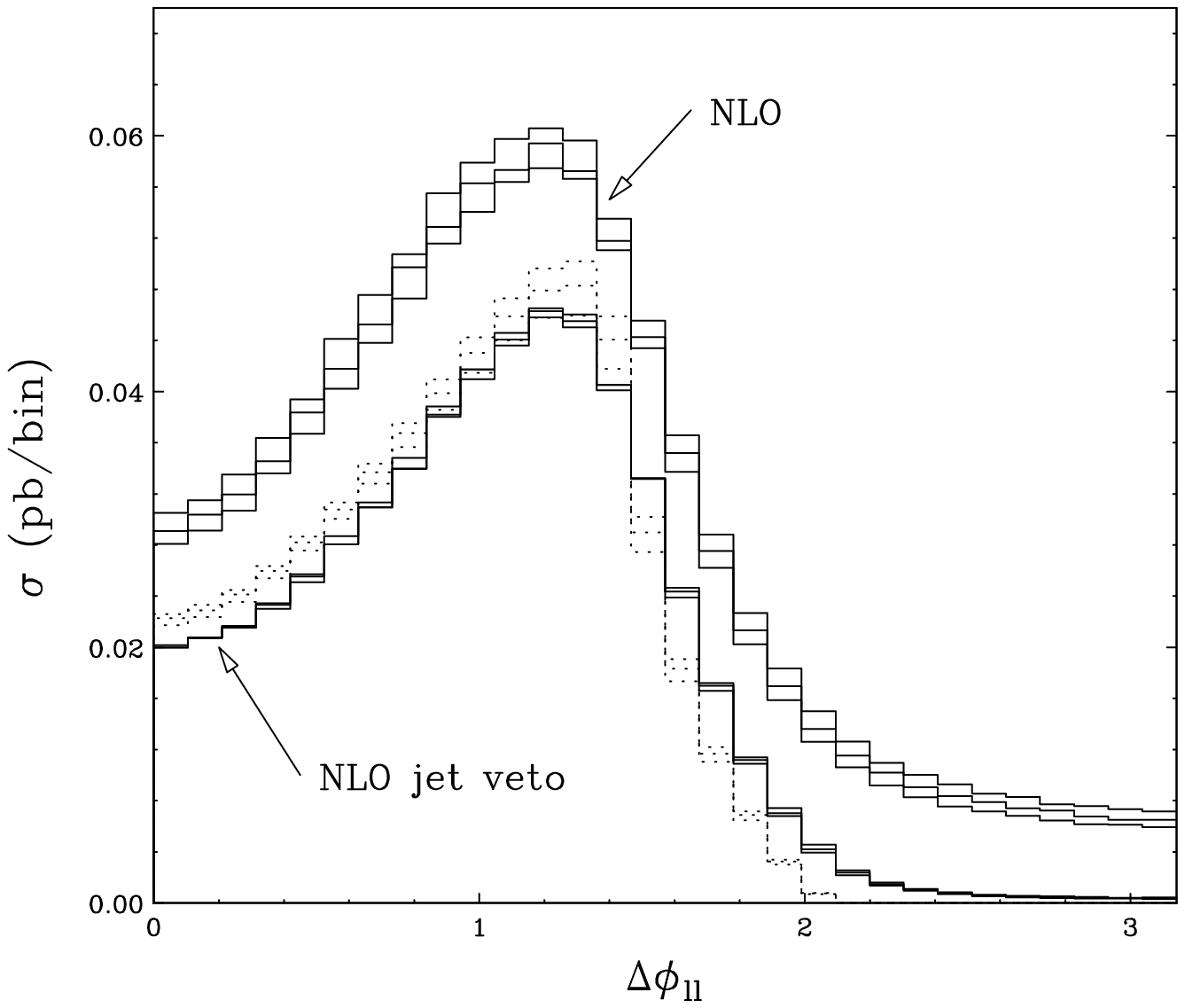,width=0.64\textwidth,clip=} }
   \ccaption{}{ \label{fig:scaleZY} Scale dependence of
   $\sigma^{(NLO)}$ without jet-veto (upper solid curves),
   $\sigma^{(NLO)}$ with jet-veto (lower solid curves) and
   $\sigma^{(LO)}$ (dotted curves). The scale has been varied
   according to $\frac{\mu_{\rm st}}{2} < \mu < 2 \mu_{\rm st}$.   }
\end{figure}

In Figure~\ref{fig:scaleZY} we study the lepton correlation in the
azimuthal angle for $Z\gamma$ production,
$\Delta\phi_{\ell\ell}=|\phi_{\ell^-}-\phi_{\ell^+}|$.  Notice that
this observable can be studied at NLO since we take fully into account
the spin correlations between the leptons in our implementation of the
one-loop corrections. The NLO corrections are rather sizeable and
increase the cross section by $50\%$ for small
$\Delta\phi_{\ell\ell}$.  The region $\Delta\phi_{\ell\ell}>2$ (with
our standard cuts) is kinematically forbidden unless a jet with a high
transverse momentum is produced. Therefore, the cross section vanishes
at LO and it is strongly suppressed for the NLO calculation with
jet-veto. In this region, the full NLO calculation is effectively only
a LO calculation and its scale dependence becomes larger, as expected.

Because there is no radiation zero appearing at LO for $Z\gamma$
production, the NLO corrections are under better control in the
kinematical region where the LO cross section does not
vanish. Nevertheless, for large transverse momentum, the process with
a $qg$ initial state again dominates the NLO contribution and the
corrections increase considerably \cite{OhnemusZ95}.

\bigskip

Finally we mention that we also considered a more stringent photon
isolation prescription, introduced by Frixione \cite{frixione}. This
prescription completely eliminates the fragmentation contribution. 
We have checked that the main features of all studied distributions
remain unchanged when it is imposed.

\section{Sensitivity to Anomalous Couplings}

The study of triple gauge boson couplings is motivated by the hope
that some physics beyond the Standard Model leads to a modification of
these couplings which eventually could be detected. In order to
quantify the effects of the new physics an effective Lagrangian is
introduced which in principle contains all Lorentz invariant
terms. The prefactor of these operators are the anomalous couplings. A
general approach is impractical, since it would lead to a
proliferation of new couplings. Therefore, some additional constraints
have to be imposed.

The usual choice for $W\gamma$ production is to insist on
electromagnetic gauge invariance and on $C$ and $P$ invariance. Also,
only operators of dimension six or less are considered. This leads to
a momentum-space vertex $W^-_\alpha(q) W^+_\beta(\qb) \gamma_\mu(p)$
(where all momenta are outgoing $p + q + \qb = 0$ ) which can be
written as \cite{HPZH87, DKS99}
\bea
 \Gamma^{\alpha \beta \mu}_{W W \gamma}(q, \qb, p) &=& 
  \qb^\alpha g^{\beta \mu} 
    \biggl( 2 + \Delta\kappa^\gamma + \lambda^\gamma {q^2\over M_W^2} \biggr) 
 - q^\beta g^{\alpha \mu}
    \biggl( 2 + \Delta\kappa^\gamma + \lambda^\gamma {\qb^2\over M_W^2}
\biggr) \nn \\  
&& \hskip 1 cm
 + \bigl( \qb^\mu - q^\mu \bigr) 
 \Biggl[ - g^{\alpha \beta} \biggl( 
   1 + {1\over2} p^2 \frac{\lambda^\gamma}{M_W^2} \biggr) 
 +\frac{\lambda^\gamma}{M_W^2} p^\alpha p^\beta \Biggr] \,,
\label{WWYvertex}
\eea
where the overall coupling has been chosen to be $-|e|$. Note that in
the Feynman rule for this vertex there is also a factor $i$ that is
conventionally not included in $\Gamma^{\alpha \beta \mu}$. In the
Standard Model we have $\dk^\gamma = \lambda^\gamma = 0$.

For $Z\gamma$ production, we consider operators up to dimension 8 (all
of them $C$ odd) resulting in $ZZ\gamma$ and $Z\gamma\gamma$
couplings. The non-standard $Z_\alpha(q_1) \gamma_\beta(q_2) Z_\mu(p)$
momentum-space vertex is given by \cite{GLR99}
\bea
 \Gamma^{\alpha \beta \mu}_{Z \gamma Z}(q_1, q_2, p) = 
   \frac{i(p^2-q_1^2)}{M_Z^2} \Biggl( 
   && h_1^Z \bigl( q_2^\mu g^{\alpha\beta} - q_2^\alpha g^{\mu \beta}
   \bigr)
   + \frac{h_2^Z}{M_Z^2} p^\alpha \Bigl( P\cdot q_2\ g^{\mu\beta} -
            q_2^\mu p^\beta \Bigr) \nn \\
   &-& h_3^Z \varepsilon^{\mu\alpha\beta\nu} q_{2\, \nu} 
   - \frac{h_4^Z}{M_Z^2} \varepsilon^{\mu\beta\nu\sigma} p^\alpha
p_\nu q_{2\, \sigma} \Biggl)
\label{ZZYvertex}
\eea where the overall coupling has been chosen to be $|e|$ (and
$\epsilon^{0123}=+1$). This vertex is absent altogether in the
Standard Model. The non-standard $Z_\alpha(q_1) \gamma_\beta(q_2)
\gamma_\mu(p)$ momentum-space vertex can be obtained from
\eqn{ZZYvertex} by setting $q_1^2 \to 0$ and replacing $h_i^Z \to
h_i^\gamma$. Notice that the vertex differs from the one implemented
in ref.~\cite{BHOZ98} by an overall factor $i$, which ensures the
hermiticity of the corresponding effective Lagrangian\footnote{There
is also a sign difference in the $h_{3,4}$ contributions coming from
the different definition of $\epsilon^{0123}$ in
ref.\cite{BHOZ98}}~\cite{GLR99}. Furthermore, the $i$ factor modifies
the interference pattern of anomalous coupling and Standard Model
amplitudes: $CP$-violating $h_{1,2}$ contributions do not interfere
with the Standard Model ones, whereas the $CP$-conserving $h_{3,4}$ do
\cite{GLR99}. Therefore, with the corrected vertex, there are
contributions linear in $h_{3,4}$ and the cross section is generally
not invariant anymore under a change of sign of $h_{3,4}$.  This must
be considered in a future precise analysis of anomalous couplings from
experimental data since the limits for $h_{3,4}$ will not be symmetric
at variance with present analyses.

The anomalous couplings spoil the gauge cancellation in the high
energy limit and, therefore, will lead to violation of unitarity for
increasing partonic center of mass energy $\sqrt{\hat{s}}$. Usually, in an
analysis of anomalous couplings from experimental data in 
hadronic collisions this problem is circumvented by supplementing the
anomalous couplings, $\alpha_{\rm AC}$, with form factors. A common
choice for the form factor is
\beq
\alpha_{\rm AC} \to \frac{\alpha_{\rm
    AC}}{(1+\frac{\hat{s}}{\Lambda^2})^n}
\label{formfac}
\eeq
where $n$ has to be large enough to ensure unitarity and $\Lambda$ is
interpreted as the scale for new physics. Obviously, this procedure is
rather ad hoc and introduces some arbitrariness \cite{mdobbs}. Also, it is not
really consistent with the effective theory approach. Increasing
anomalous contributions would require the inclusion of even higher
dimensional operators. At the end of this section section we will
address the question on how to avoid this arbitrariness in an analysis
of anomalous couplings at hadron colliders.

\bigskip
\begin{figure}[h]
\centerline{ 
   \raisebox{4.7cm}
      {\small $\frac{\sigma_{\rm AC}}{\sigma_{\rm SM}}$}
   \hspace*{1.2cm}
   \raisebox{7.5cm}{\scriptsize $\ln E_T^\ell$}
   \hspace*{4.5cm}
   \raisebox{6.5cm}{\scriptsize $\ln E_T^\gamma$}
   \hspace*{-7.5cm}
   \epsfig{figure=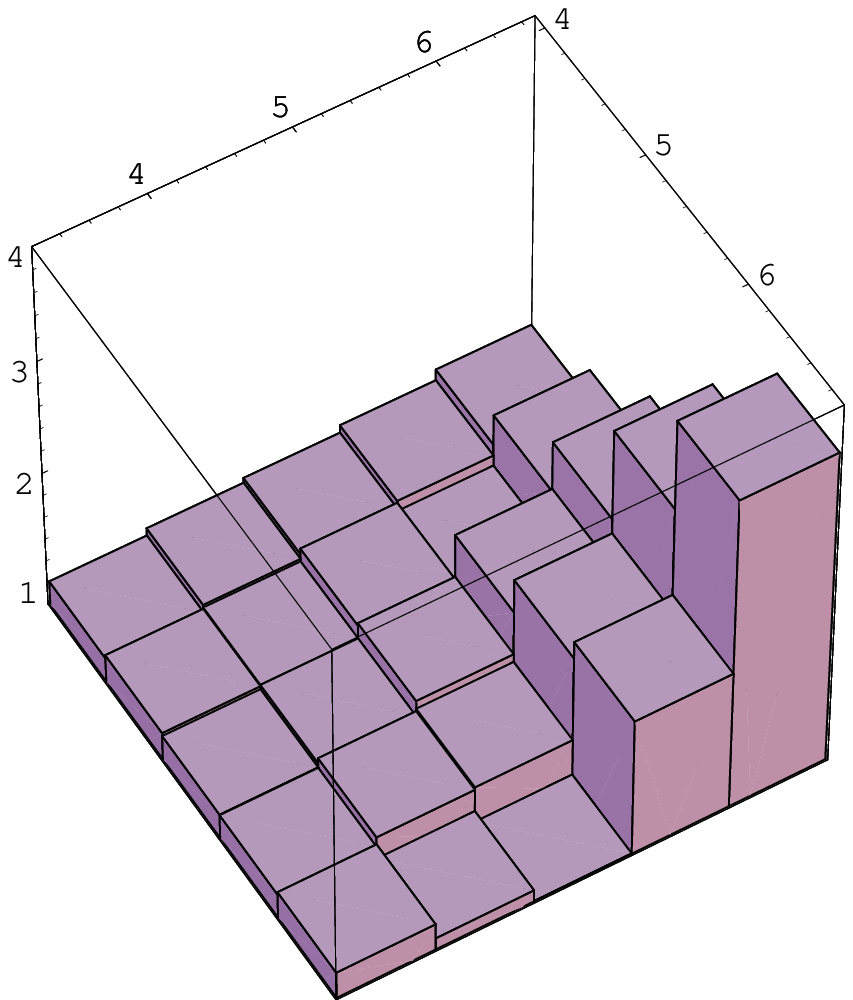,width=0.44\textwidth,clip=} }
   \ccaption{}{ \label{fig:DoubBinWY} Ratio of $\sigma_{\rm AC}$ and
   $\sigma_{\rm SM}$ at NLO for $\mu = \mu_{\rm st}$. The anomalous
   couplings in $\sigma_{\rm AC}$ have been chosen as
   $\Delta\kappa^\gamma = 0.08, \lambda^\gamma = 0.02$ and a form
   factor defined in \eqn{formfac} with $n=2$ and $\Lambda
   =$~2~TeV has been used. }
\end{figure}                                                              

Anomalous couplings mainly affect the events with large $\hat{s}$ or
large $p_T$. Since the total cross section is dominated by low $p_T$
events this is not a good observable to get tight constraints on
anomalous couplings. A more promising possibility is to consider a
double binned cross section. We therefore consider the total cross
section binned in $p_T^\ell$ and $p_T^\gamma$ for the process $pp\to
W^+\gamma \to \ell^+ \nu \gamma$ at the LHC. In
Figure~\ref{fig:DoubBinWY} we show the ratio of $\sigma_{\rm AC}$ over
$\sigma_{\rm SM}$ where $\sigma_{\rm SM}$ is the Standard Model cross
section and $\sigma_{\rm AC}$ is the cross section obtained with
$\dk=0.08$ and $\lambda=0.02$ (both within the present experimental
limits from LEP~\cite{expLEP} and Tevatron~\cite{D0WY}) and a form
factor as defined in \eqn{formfac} with $n=2$ and $\Lambda=$~2~TeV. As
expected, the ratio is large for the high $p_T$ bins, whereas it is
very close to one for the low $p_T$ bins. We checked that the
uncertainty coming from the scale variation is much smaller than the
effect of the anomalous couplings for the high $p_T$ bins.

\begin{figure}[h]
\centerline{ \epsfig{figure=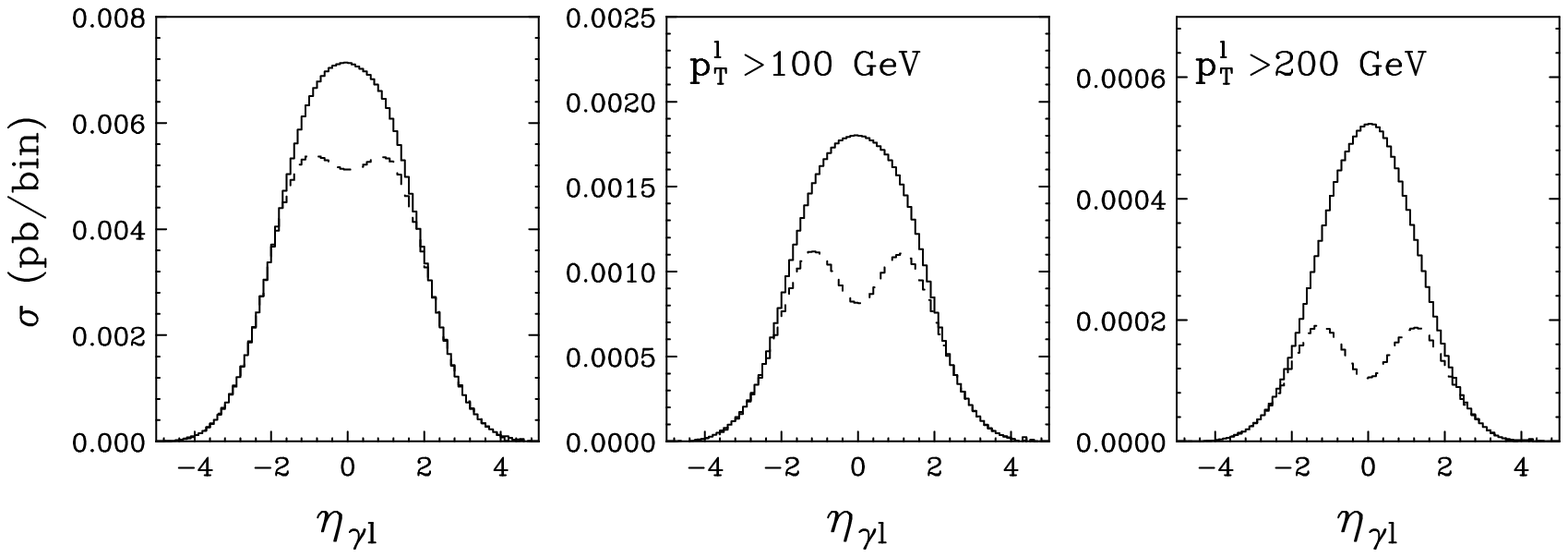,width=0.94\textwidth,clip=} }
   \ccaption{}{ \label{fig:dipEY} $\Delta\eta_{\gamma\ell}$
   distribution at NLO without (dashed curves) and with anomalous
   couplings $\Delta\kappa^\gamma = 0.08, \lambda^\gamma = 0.02$
   (solid curves). In the first plot we applied only standard cuts and
   the jet-veto, in the second plot there is an additional cut
   $p_T^\ell>100$~GeV, and in the third plot $p_T^\ell>200$~GeV.  }
\end{figure}

Another possibility to get a large effect due to anomalous couplings
is to consider the approximate radiation zeros present in the
$W\gamma$ process \cite{BMS79}. At tree-level in the Standard Model,
the $\Delta\eta_{W\gamma}\equiv \eta_W-\eta_\gamma$ distribution has a
radiation zero. This dip is filled by next-to-leading order
corrections and anomalous effects. In order to get an observable
quantity, we do not consider $\Delta\eta_{W\gamma}$ but rather
$\Delta\eta_{\gamma\ell}$ \cite{BEL94}. This will wash out the dip as
the rapidity of the lepton is not equal to the rapidity of the
$W$. However, requiring the energy (or the transverse momentum) of the
lepton to be large enough forces the lepton to follow closely the $W$
direction. Also, applying a jet-veto reduces the effects of the
next-to-leading order corrections. Thus, for larger $p_T^\ell$ even
the next-to-leading order $\Delta\eta_{\gamma\ell}$ distribution shows
a clear dip in the Standard Model. This is illustrated in
Figure~\ref{fig:dipEY}, where we show the $\Delta\eta_{\gamma\ell}$
distribution for the standard cuts and the additional cuts
$p_T^\ell>100$~GeV and $p_T^\ell>200$~GeV respectively. In all figures
we apply our jet-veto. Also shown are the three distributions with
anomalous couplings, which are chosen as for
Figure~\ref{fig:DoubBinWY}. Clearly, the effect is dramatic for high
energy leptons but of course, the disadvantage of applying such a cut
is a big loss in statistics.

\bigskip

We now turn to the question whether it is possible to avoid using form
factors in the analysis of anomalous couplings at hadron
colliders. This would bring these analyses more into line with those
at $e^+e^-$ colliders.  In order to do so one should analyze the data
at fixed values of $\hat{s}$, as it is done at LEP. This results in
limits for the anomalous parameters which are a function of
$\hat{s}$. 

\bigskip
\begin{figure}[h]
\centerline{ \epsfig{figure=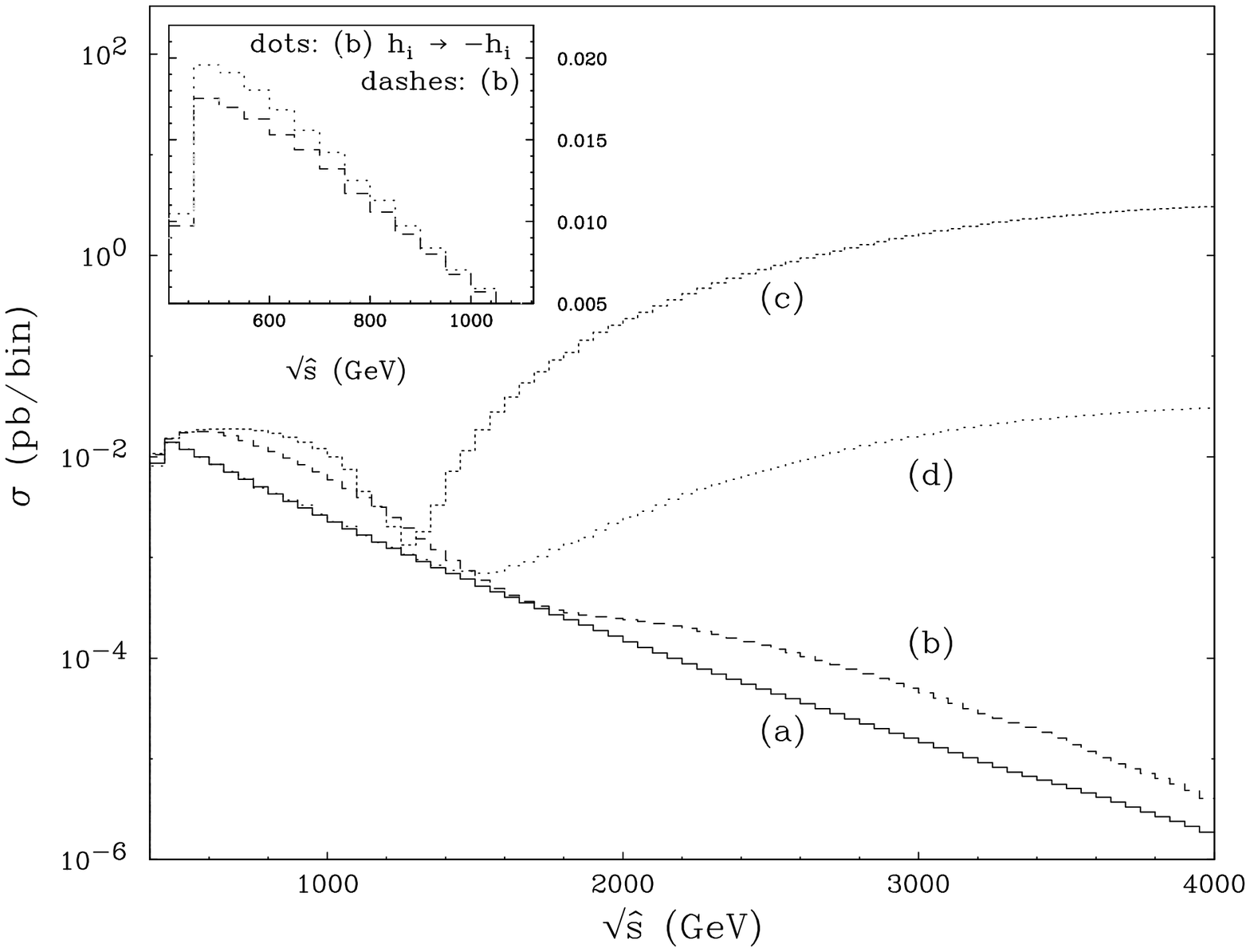,width=0.7\textwidth,clip=} }
   \ccaption{}{ \label{fig:anomZY} Cross section at NLO for $\mu=
   \mu_{\rm st}$ with standard cuts and $p_T^\gamma > 200$~GeV and
   $p_T^Z > 200$~GeV. (a) no anomalous couplings; (b)
   $h_3^\gamma=h_3^Z=0.01,\ h_4^\gamma=h_4^Z=10^{-4}$ and the usual
   dipole form factor with $\Lambda=2$~TeV; (c)
   $h_3^\gamma=h_3^Z=0.01,\ h_4^\gamma=h_4^Z=10^{-4}$ and no form
   factor; (d) $h_3^\gamma=h_3^Z=0.001,\ h_4^\gamma=h_4^Z=10^{-5}$ and
   no form factor. The inset plot shows again (b) and (b) with the opposite
   sign in the anomalous couplings. } 
\end{figure}                                                              

Obviously, it is possible to do such analysis for the production of
$Z\gamma$ when both leptons are detected. Since the center of mass
partonic energy can be reconstructed from the kinematics of the final
state particles\footnote{To simplify the discussion we assume that
{\it all} final state particles are detected, including the jets.}
the cross section can be measured for different bins of fixed
$\hat{s}$~\cite{GLR99}. As an example, we show in
Figure~\ref{fig:anomZY} the cross section as a function of
$\hat{s}$. In order to enhance the effect of the anomalous couplings
we do not only apply our standard cuts but we also require $p_T^\gamma
> 200$~GeV and $p_T^Z > 200$~GeV. We show four curves: the curve (a)
is the Standard Model result; curve (b) includes anomalous couplings
in the standard way, that is with a form factor as defined in
\eqn{formfac} with $n=3$ ($n=4$) for $h_3^V (h_4^V)$ and
$\Lambda=2$~TeV and we have set $h_3^\gamma=h_3^Z=0.01,\
h_4^\gamma=h_4^Z=10^{-4}$; curve (c) uses the same values for
$h_i^{Z/\gamma}$ but does not include any form factor; finally curve
(d) does also not include any form factors but the anomalous couplings
are smaller $h_3^\gamma=h_3^Z=0.001,\ h_4^\gamma=h_4^Z=10^{-5}$. For
all of them we set $h_1^V=h_2^V=0$.  Of course, for large $\hat{s}$
the effects are much more dramatic if the form factor is omitted and
at some point the corresponding curves would violate unitarity. This
simply reflects the breakdown of the effective theory approach in this
region. The idea behind Figure~\ref{fig:anomZY} is that such an
analysis can be done for suitably defined bins in $\hat{s}$. As a
result, for each bin, i.e. each value of $\hat{s}$ a bound on the
anomalous couplings can be obtained.

In the inset plot we present again curve (b) and the one corresponding
to the parameters of (b) but with the opposite sign for values of the
anomalous couplings $h_3^\gamma=h_3^Z=-0.01,\
h_4^\gamma=h_4^Z=-10^{-4}$.  From there the effect of the interference
between AC and SM results can be observed, i.e. the contribution of
linear terms in $h_3^V$ and $h_4^V$ appearing in the squared
amplitudes due to the correct treatment of the $i$ factor in the
$Z\gamma V$ vertex \cite{GLR99}.  For this particular configuration,
the interference effects are mostly relevant at $\sqrt{\hat{s}}<1$ TeV
and modify the cross section by more than 10\%. Clearly, they must be
taken into account in a precise extraction of anomalous couplings from
future experimental data.  Results for $h_{1,2}^V$ couplings are
similar to those obtained for the same values of $h_{3,4}^V$. The only
difference comes from the fact that the $CP$-violating couplings do
not interfere with the SM.  For a configuration like (b) with
$h_1^\gamma=h_1^Z=0.01,\ h_2^\gamma=h_2^Z=10^{-4}$, the cross section
is given by the average of both curves in the inset plot of
Figure~\ref{fig:anomZY}.

The situation is more complicated for $W\gamma$ production since the
neutrino is not observed. Nevertheless, by identifying the transverse
momentum of the neutrino with the missing transverse momentum, and
assuming the $W$ boson to be on shell, it is possible to reconstruct
the neutrino kinematics (particularly the longitudinal momentum) with
a twofold ambiguity. In the case of the Tevatron, since it is a $p\bar{p}$
collider, it is possible to choose the `correct' neutrino kinematics
73\% of the times by selecting the maximum (minimum) of the two
reconstructed values for the longitudinal momentum of the neutrino for
$W^+\gamma$ ($W^-\gamma$)~\cite{Ben95}.

\begin{figure}
\centerline{   
   \raisebox{0.5cm}{\scriptsize $\sqrt{\hat{s}}$}
   \hspace*{2cm}
   \raisebox{-0.5cm}{\scriptsize $\sqrt{\hat{s}_{\rm min}}$}
   \hspace*{2.8cm}
   \raisebox{2.7cm}{\scriptsize $\sigma$}
   \hspace*{-6.7cm}
   \epsfig{figure=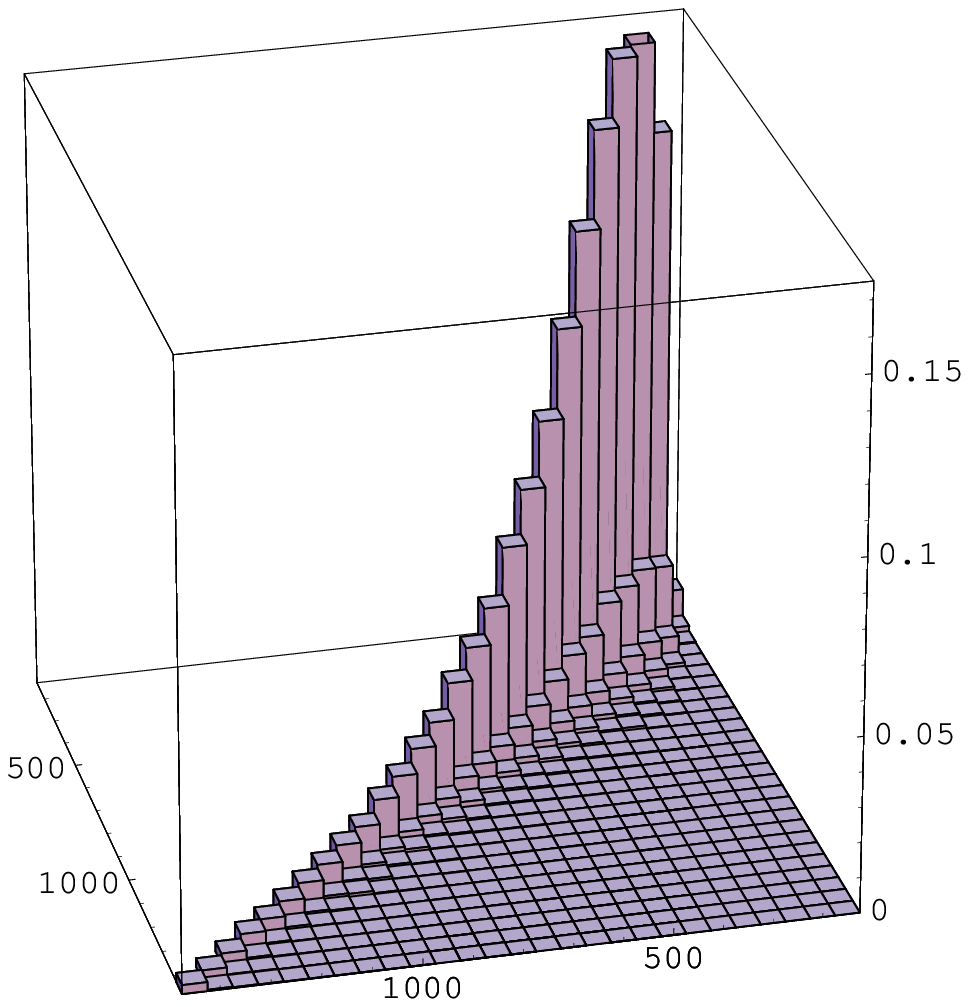,width=0.38\textwidth,clip=} \hfill
   \raisebox{0.5cm}{\scriptsize $\sqrt{\hat{s}}$}
   \hspace*{2cm}
   \raisebox{-0.5cm}{\scriptsize $\sqrt{\hat{s}_{\rm min}}$}
   \hspace*{2.8cm}
   \raisebox{2.7cm}{\scriptsize $\sigma$}
   \hspace*{-6.7cm}
   \epsfig{figure=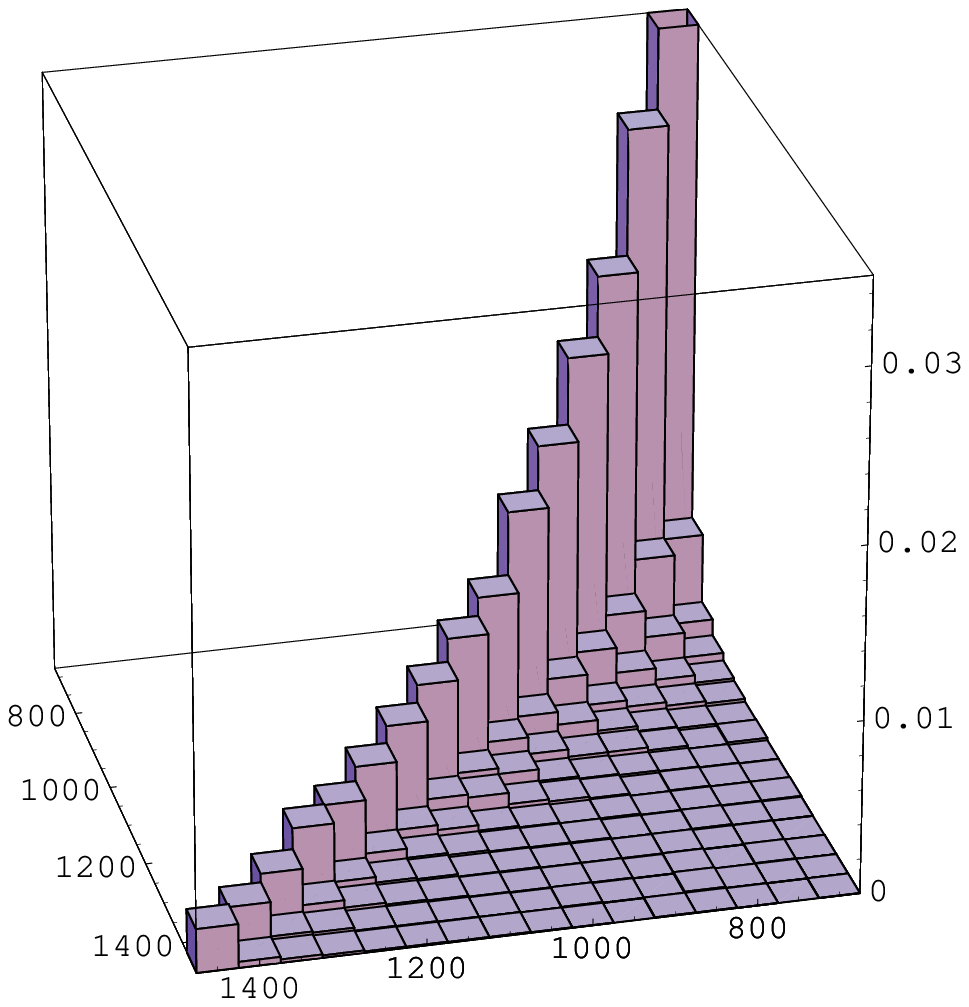,width=0.38\textwidth,clip=} }
   \ccaption{}{ \label{fig:ShatWY} The cross section for $W^+\gamma$
   production(in pb/bin) as as function of $\sqrt{\hat{s}}$ and
   $\sqrt{\hat{s}_{\rm min}}$ (in GeV) in order to illustrate the
   steep fall of $\sigma$ for increasing
   $|\sqrt{\hat{s}}-\sqrt{\hat{s}_{\rm min}}|$. The left plot shows
   the full range of $\sqrt{\hat{s}}$ whereas the right plot
   concentrates on the particularly interesting large $\sqrt{\hat{s}}$
   region. }
%
%
\bigskip
\centerline{   
   \raisebox{0.5cm}{\scriptsize $\sqrt{\hat{s}}$}
   \hspace*{2cm}
   \raisebox{-0.5cm}{\scriptsize $\sqrt{\hat{s}_{\rm min}}$}
   \hspace*{2.8cm}
   \raisebox{2.7cm}{\scriptsize $\sigma$}
   \hspace*{-6.7cm}
   \epsfig{figure=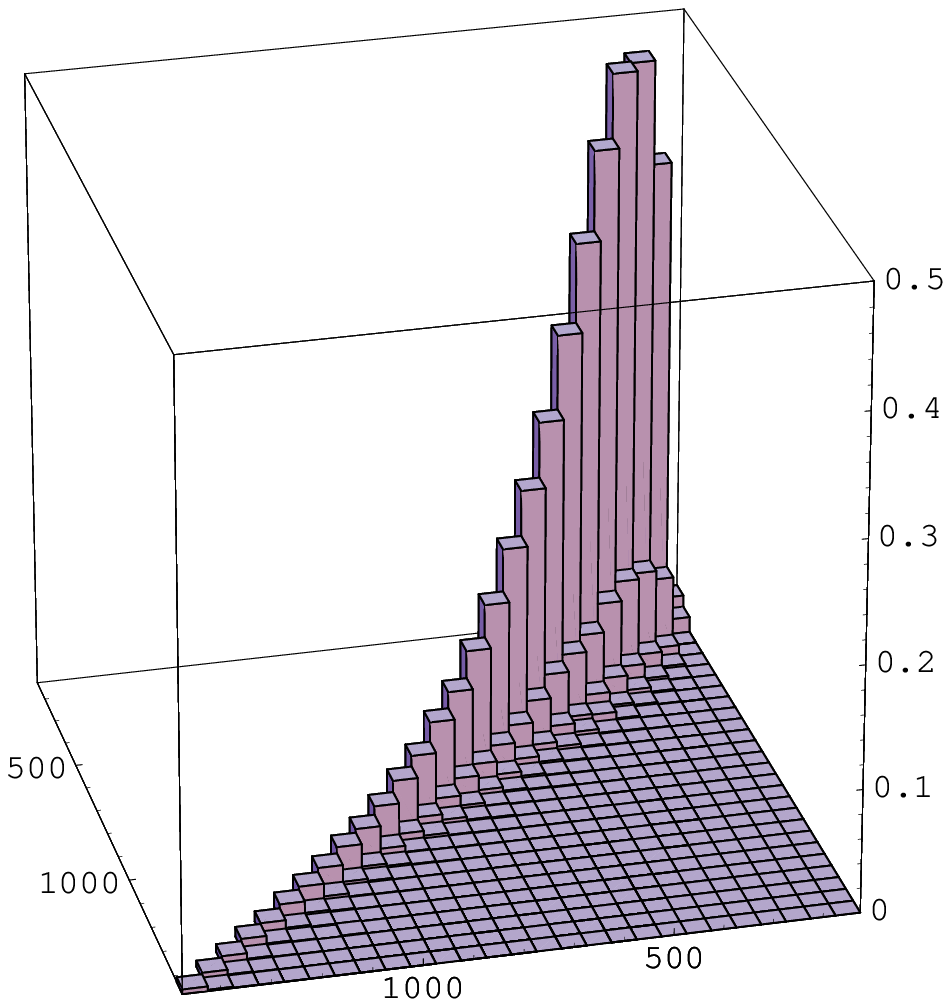,width=0.38\textwidth,clip=}
   \hfill
   \raisebox{0.5cm}{\scriptsize $\sqrt{\hat{s}}$}
   \hspace*{2cm}
   \raisebox{-0.5cm}{\scriptsize $\sqrt{\hat{s}_{\rm min}}$}
   \hspace*{2.8cm}
   \raisebox{2.7cm}{\scriptsize $\sigma$}
   \hspace*{-6.7cm}
   \epsfig{figure=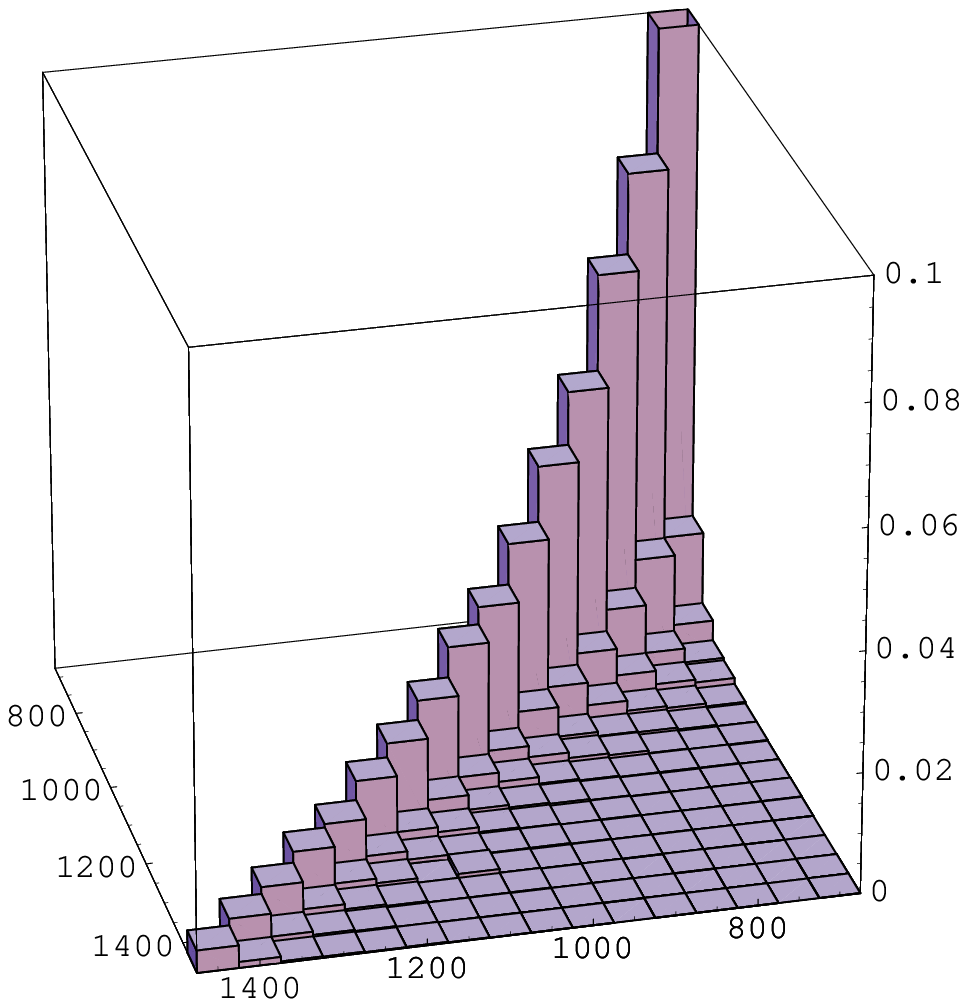,width=0.38\textwidth,clip=} }
   \ccaption{}{ \label{fig:ShatacWY} The same as
   Figure~\ref{fig:ShatWY} but with huge anomalous couplings
   $\Delta\kappa = 0.8, \ \lambda = 0.2$ and $\Lambda =$~1~TeV. } 
\end{figure}                                                              

This is not true at the LHC, where due to the symmetry of the
colliding beams both reconstructed kinematics have equal chances to be
correct. Fortunately, in the case of anomalous couplings, we are
interested in a efficient way to reconstruct the $\hat{s}$ rather than
the full kinematics. Again there are two possible values of
$\hat{s}$. It turns out that there is a simple method to choose the
`correct' one 66\% of the times at the LHC (73\% of the times at
Tevatron) by selecting the minimum $\hat{s}_{\rm min}$ of the two
reconstructed values. This applies to both $W^+\gamma$ and $W^-\gamma$
production. Furthermore, we checked that the selected value
$\hat{s}_{\rm min}$ differs in almost 90\% of the events by less than
10\% from the exact value $\hat{s}$. This is likely to be precise
enough, since the data will be collected in sizeable bins of $\hat{s}$
and the anomalous parameters are not expected to change very rapidly
as a function of the energy. To quantify the advantage of the method,
we show in Figures~\ref{fig:ShatWY} and \ref{fig:ShatacWY} the
correlations of $\sqrt{\hat{s}_{\rm min}}$ with $\sqrt{\hat{s}}$.  The
cross section drops very rapidly for increasing $\sqrt{\hat{s}}-
\sqrt{\hat{s}_{\rm min}}$.  This correlation also holds in the
particularly interesting large $\sqrt{\hat{s}}$ region and also for
the anomalous contribution. To investigate the latter point, we show
in Figure~\ref{fig:ShatacWY} the correlation for (already
experimentally ruled out) huge values of $\Delta\kappa = 0.8$ and
$\lambda = 0.2$. For this figure we still use an ordinary form factor
but in order to increase the anomalous contribution further we set
$\Lambda=1$~TeV.

\begin{figure}[h]
\centerline{
   \raisebox{0.5cm}{\scriptsize $\sqrt{\hat{s}}$}
   \hspace*{2cm}
   \raisebox{-0.5cm}{\scriptsize $\sqrt{\hat{s}_{\rm min}}$}
   \hspace*{2.8cm}
   \raisebox{2.7cm}{\scriptsize $\sigma$}
   \hspace*{-6.7cm}
   \epsfig{figure=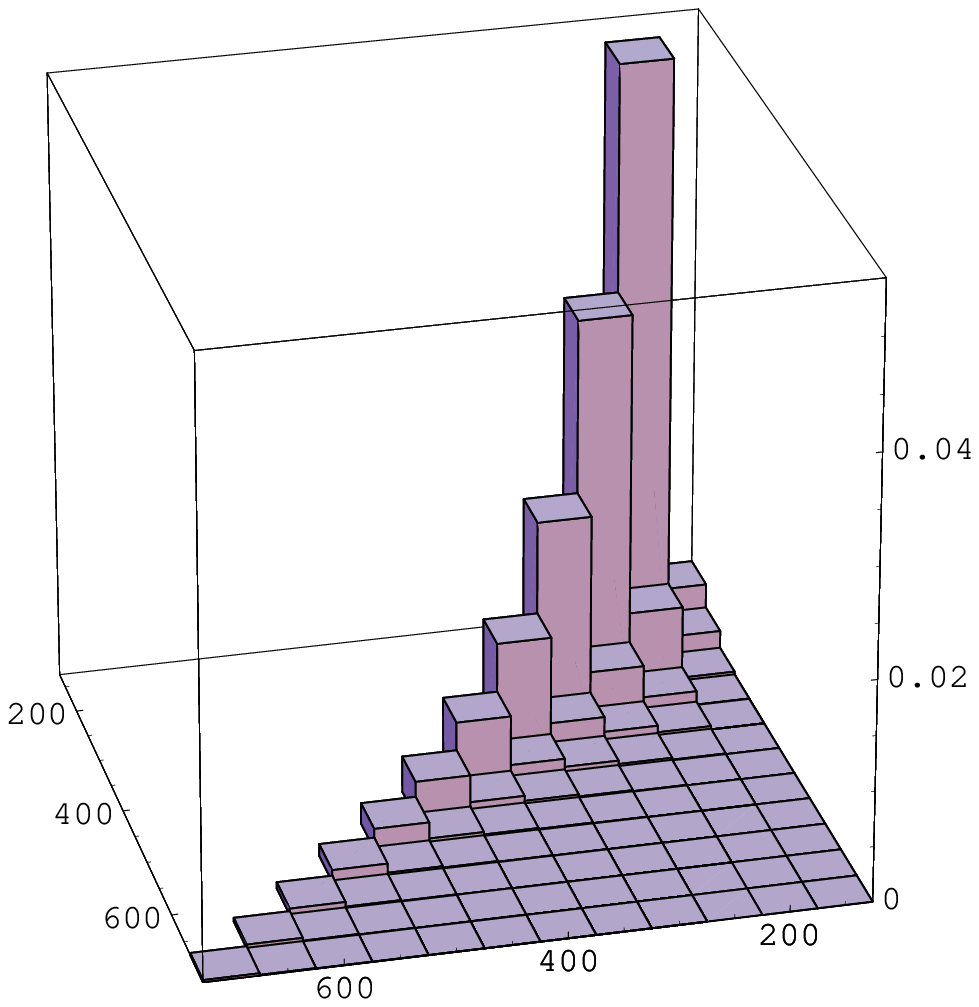,width=0.38\textwidth,clip=} \hfill
   \raisebox{0.5cm}{\scriptsize $\sqrt{\hat{s}}$}
   \hspace*{2cm}
   \raisebox{-0.5cm}{\scriptsize $\sqrt{\hat{s}_{\rm min}}$}
   \hspace*{2.8cm}
   \raisebox{2.7cm}{\scriptsize $\sigma$}
   \hspace*{-6.7cm}
   \epsfig{figure=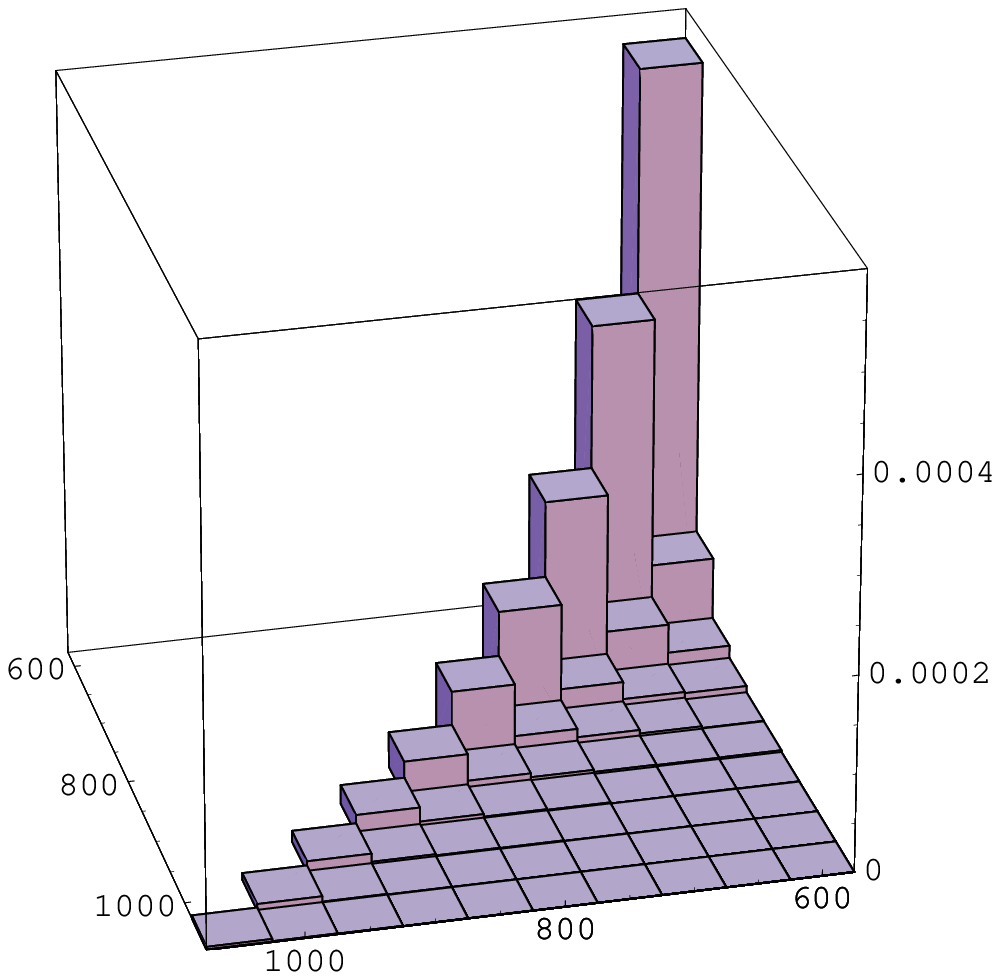,width=0.38\textwidth,clip=}
 }
   \ccaption{}{ \label{fig:ShatWYtev} The same as
   Figure~\ref{fig:ShatWY} but for Tevatron.}
\end{figure}

Finally, in Figure~\ref{fig:ShatWYtev} we show the same correlation
for $W\gamma$ production at the Tevatron (Run~II), which corresponds
to $p\bar{p}$ scattering at $\sqrt{s}=2$~TeV. We impose the same
kinematical cuts as for the LHC, with the following exceptions: for
the transverse momentum of the photon we require $p_T^\gamma > 10$~GeV
and the rapidities of the observed lepton has to be in the range
$|\eta|<1.5$ (instead of 2.4). As for the LHC, there is a strong
correlation between the 'true' center of mass energy $\sqrt{\hat{s}}$
and the 'reconstructed' one $\sqrt{\hat{s}_{\rm min}}$. Again, this
correlation also holds in the large $\sqrt{\hat{s}}$ region. In
Figure~\ref{fig:ShatWYtev} no anomalous couplings are included, but we
have checked that also for the Tevatron the inclusion of large anomalous
couplings does not spoil the correlations.

\bigskip
                                                         

As a result of these investigations we conclude that even in the case
of $W\gamma$ production reliable bounds for anomalous couplings as a
function of $\hat{s}$ can be obtained. To this end one merely has to
do an analysis as in the $Z\gamma$ case but with $\hat{s}_{\rm min}$
replacing $\hat{s}$. This has the advantage that the anomalous effects
can be quantized without introducing the ambiguity of form
factors. Such a procedure would certainly facilitate a comparison of
various bounds from different experiments. Finally, we note that in
principle any quantity which has a very strong correlation with
$\hat{s}$ can be used. However, we could not find any better candidate
than $\hat{s}_{\rm min}$. In particular, the correlations of $\hat{s}$
with the cluster mass and transverse mass respectively is not quite as
strong \cite{Durh99}.

\section{Conclusions}

In this work, we have presented a general purpose Monte Carlo program
for the calculation of any infrared safe observable in $W\gamma$ and
$Z\gamma$ production at hadron colliders at next-to-leading order in
$\alpha_s$.  The leptonic decay of the $W$ and $Z$-boson respectively
has been included in the narrow-width approximation. We retained all
spin information via decay angle correlations and thereby generalized
previous calculations~\cite{BHOW93,OhnemusZ95,BHOZ98}. We also
included anomalous triple gauge boson couplings at NLO in $\alpha_s$
and presented the analytical expressions for the corresponding amplitudes.

As an illustration of the usefulness of the program, we have studied
several observables for the LHC. Generally we find that the NLO
corrections are relevant for all of them, confirming results of
refs.~\cite{OhnemusWZ93,BHOW93,OhnemusZ95,BHOZ98}.

Moreover, we searched for the kinematical regions were the effect of
anomalous couplings is amplified and proposed an alternative way to
study its energy dependence. Using the strong correlations between the
partonic center of mass energy and a measurable variable,
$\hat{s}_{\rm min}$, makes it possible to extract anomalous couplings
from the data without need to introduce ad hoc form factors. Such an
analysis is possible even for the $W\gamma$ process and, in our view,
should be undertaken at the Tevatron and LHC.

\section*{Acknowledgments}

It is a pleasure to thank Z. Kunszt for his participation on the
 initial stages  of  this work and for helpful discussions.
D.~de F. would like to thank the Department of Physics of the
University of Durham for its hospitality while part of this work was
carried out.

This work was partly supported by the EU Fourth Framework Programme
`Training and Mobility of Researchers', Network `Quantum
Chromodynamics and the Deep Structure of Elementary Particles',
contract FMRX-CT98-0194 (DG 12 - MIHT).

\section{Appendix}

The helicity amplitudes that are needed for the calculation of
$W\gamma$ and $Z\gamma$ production at next-to-leading order can be
found in ref.~\cite{DKS98}. In this appendix we list the additional
amplitudes that are needed in the presence of anomalous couplings
which lead to the non-standard vertices as given in
eqs.~(\ref{WWYvertex}) and (\ref{ZZYvertex}).  We use the notation and
conventions of refs.~\cite{DKS98,DKS99}.

We start with the $W\gamma$ amplitudes. In order to maintain
electromagnetic gauge invariance we set $g_1^\gamma = 0$ and only
allow anomalous couplings $\dk^\gamma$ and $\lambda^\gamma$. The only
diagram that gets modified by these anomalous couplings is the diagram
with a $WW\gamma$ coupling. Fortunately, this diagram does not
contribute to the rather complicated finite pieces $F^a_\gamma,
F^b_\gamma$ of the amplitudes, given in eqs.~(4.6) and (4.7) of
ref.~\cite{DKS98}. Therefore, only the tree-level amplitudes have to
be computed with anomalous couplings.

We do not have to list the explicit results for all possible
helicities of the photon and the gluon. In order to reverse the
helicities of all gauge bosons, i.e. the photon and the gluon (if the
latter is present), we merely have to apply a `flip' operation, defined
as
\beq
{\rm flip2}: 1\lra 2; \ 3\lra 4; \ \spa{a}.{b} \lra \spb{a}.{b};\,\,\,
{\rm and} \, 
A^a \lra A^b \,({\rm for} \, W\gamma)
\label{flip2}
\eeq

The amplitudes $A^{\rm tree,a}_\gamma, A^{\rm tree,b}_\gamma$ given in
eqs.~(4.4) and (4.5) of ref.~\cite{DKS98} are modified as follows in
the presence of anomalous couplings:
\bea
A^{\rm tree,a}_{{\rm AC}, \gamma} &=& 
   A^{\rm tree,a}_\gamma + A^5_{{\rm AC}} \\
A^{\rm tree,b}_{{\rm AC}, \gamma} &=& 
   A^{\rm tree,b}_\gamma - A^5_{{\rm AC}} 
\eea
where
\beq
A^5_{{\rm AC}} = \frac{-i\spb4.5}{2 s_{34} (s_{12}-s_{34}) \spb3.4}
\Big( (\dk^\gamma+\lambda^\gamma) \spa1.3\spb2.5\spb3.4 +
\lambda^\gamma \spaa1.{5}.2\spb4.5 \Big) 
\eeq

As usual, this will also lead to a modification of the one-loop
amplitudes. The corresponding divergent pieces now read $\cg V A^{\rm
tree,a}_{{\rm AC}, \gamma}$ and $\cg V A^{\rm tree,b}_{{\rm AC},
\gamma}$ respectively, where
\beq
V = 
  - {1\over \eps^2} \left( {\mu^2 \over -s_{12}}\right) ^\eps 
  - {3\over 2 \eps} \left( {\mu^2 \over -s_{12}}\right) ^\eps 
  - {7\over 2} \,. 
\label{Vab}
\eeq

In the case of the bremsstrahlung amplitudes we have to consider two
cases. The additional gluon can have positive or negative helicity.
The corresponding Standard Model amplitudes are given in eqs.~(4.9) to
(4.12) of ref.~\cite{DKS98}. In the presence of anomalous couplings
$\dk$ and $\lambda$ they are modified as follows.
\bea
A^{\rm tree,a}_{{\rm AC}, 6,\gamma}(1^-,2^+,3^-,4^+,5^+,6^+) &=& 
   A^{\rm tree,a}_{6,\gamma}(1^-,2^+,3^-,4^+,5^+,6^+) + 
   A^{6 +}_{{\rm AC}} \\
A^{\rm tree,a}_{{\rm AC}, 6,\gamma}(1^-,2^+,3^-,4^+,5^-,6^+) &=& 
   A^{\rm tree,a}_{6,\gamma}(1^-,2^+,3^-,4^+,5^-,6^+) + 
   A^{6 -}_{{\rm AC}} \\
A^{\rm tree,b}_{{\rm AC}, 6,\gamma}(1^-,2^+,3^-,4^+,5^+,6^+) &=& 
   A^{\rm tree,b}_{6,\gamma}(1^-,2^+,3^-,4^+,5^+,6^+) - 
   A^{6 +}_{{\rm AC}} \\
A^{\rm tree,b}_{{\rm AC}, 6,\gamma}(1^-,2^+,3^-,4^+,5^-,6^+) &=& 
   A^{\rm tree,b}_{6,\gamma}(1^-,2^+,3^-,4^+,5^-,6^+) - 
   A^{6 -}_{{\rm AC}} 
\eea
where we defined
\bea
A^{6 +}_{{\rm AC}} &\equiv& i \frac{\spb4.5 \spaa1.{2+6}.5}{2\
s_{34}^2 (t_{126}-s_{34}) \spa1.6 \spa 6.2} \Big( \lambda^\gamma
\spa3.4\spa1.5\spa4.5 + (\Delta\kappa^\gamma + \lambda^\gamma) \spa1.3
s_{34} \Big) 
\\
A^{6 -}_{{\rm AC}} &\equiv& i \frac{\spa3.5\spa1.5}{2\
s_{34}^2 (t_{126}-s_{34}) \spa1.6 \spa 6.2} \times \nn \\
&& \qquad \qquad \Big( \lambda^\gamma
\spa3.5\spaa1.{2+6}.5\spb3.4 - (\Delta\kappa^\gamma + \lambda^\gamma)
\spaa1.{2+6}.4 s_{34} \Big)
\eea

We now turn to the amplitudes for $Z\gamma$ production. In the
Standard Model, there are no diagrams with a triple gauge boson
coupling and the corresponding amplitudes $A^{\rm tree,s}_\gamma$ can
simply be obtained as the symmetric combination of the $W\gamma$
amplitudes
\beq
A^{\rm tree,s}_\gamma = A^{\rm tree,a}_\gamma +  A^{\rm tree,b}_\gamma 
\eeq
(see eq.~(4.15) of ref.~\cite{DKS98}). These combinations can be
simplified somewhat but we refrain from listing the simplified
versions.

The amplitudes related to an anomalous $ZZ\gamma$ or $Z\gamma\gamma$
coupling will be denoted by $ A_{5/6, {\rm AC}}^{(Z\gamma)}$. In the
former case, the intermediate vector boson is a $Z$-boson whereas in
the latter case it is a $\gamma$. This results in different couplings
of the intermediate vector boson to the initial state quarks. Apart
from this difference, the amplitudes with an intermediate $Z$ and
$\gamma$ are the same. The anomalous couplings always appear in the
combination 
\beq
\h_1^{Z/\gamma} \equiv \frac{h_1^{Z/\gamma}}{M_Z^2}; \ \
\h_2^{Z/\gamma} \equiv \frac{h_2^{Z/\gamma}}{M_Z^4}; \ \
\h_3^{Z/\gamma} \equiv \frac{h_3^{Z/\gamma}}{M_Z^2}; \ \
\h_4^{Z/\gamma} \equiv \frac{h_4^{Z/\gamma}}{M_Z^4}. \
\eeq
We start with the tree-level amplitude for a positive helicity photon
\beq
A_{5, {\rm AC}}^{(Z\gamma)} = \frac{i}{4 s_{34}} \Big(
 (i \h_1^{Z/\gamma}+ \h_3^{Z/\gamma}) 2 \spa1.3\spb2.5\spb4.5 +
 (i \h_2^{Z/\gamma}+ \h_4^{Z/\gamma}) \spa1.2\spaa3.5.4\spb2.5^2 \Big)
\eeq

In order to obtain the amplitudes for a negative helicity photon we
have to apply the `flip2' operation defined in \eqn{flip2} in
addition to
\beq
\h_1^{Z/\gamma}\to -\h_1^{Z/\gamma}; \ \h_2^{Z/\gamma}\to -\h_2^{Z/\gamma};
\eeq

The anomalous bremsstrahlung amplitude with an additional positive
helicity gluon reads
\bea
A_{6, {\rm AC}}^{(Z\gamma)} &=& \frac{i}{4
s_{34}\spa1.6\spa2.6}\times \\ 
&&  \Big(
(i \h_1^{Z/\gamma}+ \h_3^{Z/\gamma}) 2 \spa1.3\spb4.5\spaa1.{2+6}.5 +
(i \h_2^{Z/\gamma}+ \h_4^{Z/\gamma}) \spaa3.5.4\spaa1.{2+6}.5^2 \Big)
\nn 
\eea
whereas for a negative helicity gluon
\bea
A_{6, {\rm AC}}^{(Z\gamma)} &=&  \frac{i}{4 s_{34}\spb1.6\spb2.6}
\times \\ 
&&  \Big(
(i \h_1^{Z/\gamma}+ \h_3^{Z/\gamma}) 2 \spb4.5\spb2.5\spaa3.{1+6}.2 +
(i \h_2^{Z/\gamma}+ \h_4^{Z/\gamma}) t_{126} \spaa3.5.4 \spb2.5^2
\Big) \nn 
\eea
Again, the operation `flip2' reverses the helicities of the photon and
the gluon. Finally we mention that in order to get the amplitudes with
a positive helicity lepton, $3^+$, we simply have to exchange $3
\leftrightarrow 4$ in the amplitudes presented above. Correspondingly, the
amplitudes with opposite helicities for the partons are obtained by a
simple  $1 \leftrightarrow 2$ crossing.

\def\np#1#2#3  {{\it Nucl. Phys. }{\bf #1} (19#3) #2}
\def\nc#1#2#3  {{\it Nuovo. Cim. }{\bf #1} (19#3) #2}
\def\pl#1#2#3  {{\it Phys. Lett. }{\bf #1} (19#3) #2}
\def\pr#1#2#3  {{\it Phys. Rev. }{\bf #1} (19#3) #2}
\def\prl#1#2#3  {{\it Phys. Rev. Lett. }{\bf #1} (19#3) #2}
\def\prep#1#2#3 {{\it Phys. Rep. }{\bf #1} (19#3) #2}
\def\zp#1#2#3  {{\it Z. Phys. }{\bf #1} (19#3) #2}
\def\rmp#1#2#3  {{\it Rev. Mod. Phys. }{\bf #1} (19#3) #2}
\def\mpl#1#2#3 {{\it Mod. Phys. Lett. }{\bf #1} (19#3) #2}


\end{document}